%Paper: hep-th/9409055
%From: Charilaos Aneziris <aneziris@hades.ifh.de>
%Date: Sat, 10 Sep 1994 21:12:39 +0200 (METDST)

\magnification=\magstep1
\nopagenumbers
\parskip 0pt
\baselineskip 16pt
\bigskip
\centerline  {\bf  Towards a Classification of Knots}
\bigskip
\centerline {Charilaos Aneziris}
\centerline {\it Institut f\"ur Hochenergienphysik Zeuthen}
\centerline {\it Platanenallee 6}
\centerline {\it 15738 Zeuthen}
\bigskip
\centerline {\bf Abstract}
\bigskip
\par We discuss the possibility of the existence of finite algorithms
that may give distinct knot classes. In particular we present two
attempts for such algorithms which seem promising, one based on knot
projections on a plane, the other on placing knots on a cubic lattice.
\bigskip
\centerline {\bf 1. Introduction}
\bigskip
\par Knot Theory has been one of the newest area of interest in
Mathematical Physics during the last two decades. In High Energy
Theory, in particular, its relevance is due to the fact that the
observables in Topological Field Theories are integrals along
distinct knot classes whose values are given by the Jones polynomials $^{(1)}$.
While a number of interesting theorems and results have been derived
in knot theory during recent years, we have yet to come up with some
procedure that would yield the distinct knot classes, which are our
object of interest.
\smallskip
\par After providing some basic definitions in knot theory we shall
present here two such attempts which may eventually yield finite
algorithms.
\smallskip
\par The first attempt is based on projecting knots on a plane and
providing such projections with ``names". Reidemeister moves may prove
the equivalence of such projections, while knot group theory can show
the inequivalence; unfortunately none of these two methods is finite,
in the sense that we have yet to establish an upper limit after which
at least one of them would terminate.
\smallskip
\par The second attempt does not depend on any knot projection; while
keeping the knot in three dimensions, we make its segments run parallel
to a system of three orthogonal axes. In this way, the sequence of these
segments provides the knot with a finite name. Here there are moves that
may show knot equivalence, and indeed the corresponding algorithm is
much simpler to be written, although needs more time to run. This
procedure is also not finite, and in addition we have yet to come up with
an inequivalence proof.
\bigskip
\centerline {{\bf 2. Definitions} $^{(2)}$}
\bigskip
\par A {\it knot} is defined as a continuous one-to-one mapping from
$S^1$ to $S^3$.
\smallskip
\par According to the definition above, a knot may not possess ``double"
points, or equivalently, no two points of $S^1$ may be projected on the
same point of $S^3$. More general, however, the target space need not be
$S^3$ but any manifold $\cal M$. For spaces other than $S^3$ or $R^3$ we
already know the set of knot classes, which in most cases is trivial, and
this is the reason why we shall restrict ourselves in ${\cal M}=S^3$. (For
${\cal M}=R^3$ we get the same result as in $S^3$, as one may easily
notice).
\smallskip
\par Two knots are defined as {\it ambient isotopic} and will be considered
as belonging to the same (equivalence) class if there is a continuous
series of knots connecting them.
\smallskip
\par If $f_1(s)$ and $f_2(s)$ are the two continuous one-to-one functions
from $S^1$ to $S^3$ defining two knots. In order for these knots to be
ambient isotopic, there must exist a function $g(s,t)$ from $S^1 \times
[0,1]$ to $S^3$ , continuous with respect to both $s$ and $t$, such that
$$ s_1 \neq s_2 \Rightarrow g(s_1,t) \neq g(s_2,t)$$
and
$$ g(s,0)=f_1(s) \wedge g(s,1)=f_2(s)$$
\smallskip
\par It is clear from the definitions above that ambient isotopy is an
equivalence relation, dividing the set of knots into classes such that all
elements of a class are ambient isotopic to each other while none is ambient
isotopic to any element of any other class. What we shall do from now
on is provide knots with numerical ``names", establish relations that
show which such names are equivalent and which not, give a criterion to
keep just one name out of each class and finally try to devise an
algorithm that shall list distinct knot classes through the names of their
representatives. While all steps mentioned above are possible, the main
hurdle we shall have to overcome is the finiteness of the algorithm, which
is due to the fact that no upper limit is known for the number of steps
needed to show the equivalence or inequivalence of two names.
\smallskip
\par One final remark we would like to make in this section before moving on
is that while knots with inverse orientation as well as knots that are mirror
images of each other, are not necessarily ambient isotopic, they are going
to be considered equivalent here. The problem of chirality and invertibility
is not going to be discussed here.
\bigskip
\centerline {\bf 3. Knot Names}
\bigskip
\par Let $K$ be a knot and $P(K)$ be some projection of $K$ on some plane $P$
such that no more than two points of $K$ are projected on the same point of
$P$. In addition, we impose the constraint that if $s_1$ and $s_2$ are two
points of $K$ that are projected on the same point $p \in P$ and if $K$ is
defined by some function $f(s)$, $f'(s_1)-f'(s_2)$ should not be
perpendicular to $P$. We now choose some point $O \in P(K)$ and one out of
the two possible directions, and move along $P(K)$. As we pass by, we
assign successive natural numbers to the ``double" points we meet, starting
from $1$ and ending to some even number $2n$. Each double point is
assigned two such numbers, one for the overcrossing and one for the
undercrossing. Let $i$ be the number for the overcrossing and $j$ the
number for the undercrossing. We then assign to this double point the pair
$(i,j)$. We now define as {\it name} of the projection $P(K)$ the set of
all such pairs. As an example, the name of the {\it trefoil} is
$\{ (1,4), (3,6), (5,2) \}$, while of the {\it figure eight} knot is
$\{ (1,4), (3,6), (5,8), (7,2) \}$ $^{(3)}$.
\smallskip
\par While this naming is going to help us while we work on our first
algorithm in section 4, in section 5 we shall use another algorithm based
on placing knots on a 3-dimensional cubic lattice on $Z^3$. The naming
procedure now goes as follows.
\smallskip
\par Let $(0,0,0)$ be our starting point, that is let $f(0)=(0,0,0)$. (If
that is not so, we can simply move the origin). We shall replace now the
knot $K$ by an equivalent knot $K'$ satisfying the following properties.
$$a) \hskip 1 cm \exists n \in N : f({k \pi \over n}) \in Z^3
\hskip 0.3 cm \forall k \in
\{ 1,2,...,2n-1 \}$$
$$b) \hskip 2cm |f({(k+1) \pi \over n}) - f({k \pi \over n})|=1$$
and
$$c) \hskip 0.5 cm f({k (\pi + x) \over n}) = x f ({k \pi \over n})
+ (1-x) f({(k+1) \pi \over n}) \hskip 0.3 cm \forall x \in (0,1)$$
\smallskip
\par We now assign $K'$ with a ``name" which is a sequence of $2n$ numbers
$a_1,a_2,...,a_{2n}$, where $a_i \in \{1,2,3,4,5,6\} \forall i \in
\{ 1,2,3,...,2n\}$ and $a_i=1,2,3,4,5,6$ means that $f( {(k+1) \pi \over n} )
-f( {k \pi \over n} )$ is equal respectively to the unit vectors along the
axes $x,y,z,-z,-y,-x$ $^{(4)}$.
\smallskip
\par Whether we use the first, the second or any other notation, once we
have established some procedure to get a sequence of such names there are
three important criteria we have to use in order to see if we got an
algorithm that yields distinct knot classes. They are the following.
\smallskip
\par a) Are there any knot classes that are missing?
\par b) Are there any classes repeated more than once?
\par c) Do all the ``names" we get through our procedure, correspond to
actual knot classes?
\bigskip
\centerline {\bf 4. The Projection Algorithm}
\bigskip
\par Let $n \in N$, $N_{2n}=\{1,2,...,2n\}$, and $S(n)$ be a set consisting
of pairs $(i,j)$ where both $i$ and $j$ belong to $N_{2n}$ and all elements of
$N_{2n}$ belong to one and only one pair of $S(n)$. One may easily show that
for a given $n$ there is a total of ${ (2n)! \over n!}$ possible such sets
$S(n)$. If we write down all $S(k)$ $\forall k \in N_n$ and try to use it as
a knot generating algorithm, we can be sure that no knot classes shall be
missing, at least no knot classes whose complexity exceeds a value
characterized by the number of double points $n$. Most classes however will
be repeated a number of times, while many sets generated will not yield any
actual knot class.
\smallskip
\par The last point can be easily shown by taking the set $\{ (1,3),(2,4) \}$.
One may easily see that it is not possible to make such a projection.
Indeed one may prove by using the {\it Jordan Curve Theorem} that no name
corresponds to a knot projection unless all its pairs consist of one odd and
one even number. This constraint in fact simplifies the writing of an
algorithm; it also reduces the possible combinations to $2^n n!$.
\smallskip
\par Even this condition, however, is sufficient, as one may see by
attempting to get the projection $\{ (1,4), (3,10), (5,8), (7,2), (9,6) \}$.
Once again for such a name the Jordan Curve Theorem is violated, but on a
different scale. Fortunately it is possible through a finite process to
eliminate all such names that violate the Theorem, while the rest will
yield actual knot classes.
\smallskip
\par All we need to do now is eliminate classes that are repeated more than
once. Knot classes may be repeated for two reasons: either we use a different
starting point or orientation, in which case we get the same projection but
under a different name, or we may get two different projections of the same
knot class, which are connected to each other through a sequence of
{\it Reidemeister} moves $^{(5)}$.
\smallskip
\par The first problem can easily be solved, since there are at most $4n$
different possibilities, $2n$ corresponding to different starting points
and $2$ possible orientations. All we need to do is establish some criterion
that will choose just one of these  possibilities and reject the others. The
criterion usually chosen is the {\it lexicographical} one.
\smallskip
\par Unfortunately eliminating equivalent projections is not that easy, the
reason being that one may not always be able to prove the equivalence of two
such projections. While Reidemeister showed the three kinds of moves that
connect two equivalent projections, and which essentially say that pairs
$(i,i\pm 1)$ and $(i,j),(i\pm 1,j\pm 1)$ are redundant and a triad of pairs
$(i,j),(i',k),(j',k')$ is equivalent to the triad $(i,k'),(i',j'),(j,k)$
where $|i'-i|=|j'-j|=|k'-k|=1$, one
does not know in advance how many moves are needed at least to connect
equivalent projections. No matter how ``hard" we try to prove that two
projections are equivalent, failing to do so does not prove that these
projections are inequivalent.
\smallskip
\par It is possible, however, to prove that two projections are inequivalent,
by showing that their {\it knot groups}, that is the fundamental groups of
their complements, are distinct. Unfortunately proving the distinctness of
two groups is not always simple; two groups may possess what on first sight
may seem completely different generators and defining relations, but in
reality be the same. A simple example of such a case is a free group
generated by $a$ and a group generated by $a$ and $b$ where $b=a^2$. One may
show the distinctness of two knot groups by assigning their generators to
elements of a known group; if such an assignment is possible for one knot
group but results in a contradiction for the other knot group, the two
knots are definitely inequivalent. This in fact was the method through
which Reidemeister proved that the trefoil is non-trivial, since its
generators can be assigned to three distinct elements of the permutation
group $S_3$, something which is impossible for the trivial knot.
\smallskip
\par This procedure is unfortunately also not finite, so the best one may
hope in order to compare two projections is to run simultaneously two
algorithms, one in order to show possible equivalence, another to show
possible inequivalence, and wait until one algorithm yields a definite
result. In order to get knot classes, one may first attempt to eliminate as
many knot classes as possible by showing that the others are equivalent to
the remaining ones, and then try to show that the remaining are inequivalent
to each other.
\bigskip
\centerline {\bf 5. The Lattice Algorithm}
\bigskip
\par Using the second algorithm as described in section 3 has a certain
advantage: it is easier to be written down. First, the parameters $a_i$ may
take any
value from $1$ to $6$ and they don't necessarily have to be different from
each other. Second, no Jordan Curve Theorem is needed to eliminate
``impossible" names; all one needs is to make sure that there are as many
$1$'s, $2$'s and $3$'s as $6$'s, $5$'s and $4$'s respectively, so that the
knot closes, and that such a closure does not occur before the end, so that no
double points exist. Third, the equivalence moves are also simpler than the
Reidemeister ones; these correspond to inverting the order of two subsequent
terms so that $a_1...a_{i-1}a_ia_{i+1}a_{i+2}...a_{2n}$ becomes
$a_1...a_{i-1}a_{i+1}a_ia_{i+2}...a_{2n}$, and to adding two opposing unit
 vectors
just before and after an existing one, so that
$a_1...a_{i-1}a_ia_{i+1}...a_{2n}$ becomes
$a_1...a_{i-1} \alpha a_i \beta a_{i+1}...a_{2n}$ where $\alpha,\beta \in
\{1,2,3,4,5,6\} \wedge \alpha + \beta =7$. One must take care of course that
no double points exist in the new knot in order for the move to be allowed.
\smallskip
\par While the algorithm is much simpler to be written down, however, it needs
much more time to run, the reason being that for a certain knot class the
value of $n$ is much higher. For the trefoil, for instance, while the
projection algorithm gave it for $n=3$, since when projected on a plane it
possesses three (at least) double points, the lattice algorithm will give it
for $n=12$, since the corresponding knot must have a length of at least $24$
units $^{(6)}$.
\smallskip
\par As far as the three criteria of section 3 are concerned, the results
come out the same as with the projection algorithm; no class is missing, all
names correspond to actual knot classes, but classes do get repeated and there
is no (finite) way to ensure that repeated classes are eliminated.
Unfortunately
there is no method similar to the knot group one that can show if two knots are
inequivalent.
\bigskip
\centerline {\bf 6. Conclusion}
\bigskip
\par While the problem of the ennumeration of knots has yet to be solved,
there is a number of promising avenues which we hope that will eventually
provide us with the final word. Here we discussed two ideas that might lead
to the establishment of a desirable algorithm; our failure to figure out a
maximum number of steps that these methods need in order to work is the only
obstacle remaining to be overcome.
\bigskip \centerline {R E F E R E N C E S} \bigskip
\par 1) E. Witten, {\it Comm. Math. Phys.} {\bf 121} 351-399.
\par 2) For more details on the subject, see for example Rolfsen, D.,
(1976) {\it Knots and Links}, (Berkeley, CA: Publish or Perish, Inc.) and
Burde G., Zieschang H., (1985) {\it Knots} (Berlin: de Gruyter).
\par 3) A similar idea has been studied by M.B. Thistlethwaite, {\it Aspects
of Topology in Memory of Hugh Dowker}, L.M.S. Lecture Notes No 93 (Cambridge
University Press 1985) 1-76.
\par 4) See also N. Madras and G. Slade (1993) {\it The Self-Avoiding Walk}
pp.276-278 and references therein.
\par 5) Reidemeister, K. (1932) {\it Knotentheorie} Ergebn. Math. Grenzgeb.,
Bd. {\bf 1}; Berlin: Springer-Verlag.
\par 6) Y.A. Diao, {\it Journal of Knot Theory and its Ramifications} {\bf 2}
413-427.
\end